\documentclass[aps,prl,twocolumn,superscriptaddress,showpacs,floatfix,altaffilletter,longbibliography]{revtex4-2}

\usepackage{graphicx}
\usepackage{grffile}
\usepackage{bm}
\usepackage{color}
\usepackage[dvipsnames]{xcolor}
\usepackage{amsmath,amssymb,amsfonts}
\usepackage{epsfig}
\usepackage{times}
\usepackage[colorlinks,bookmarks=false,citecolor=blue,linkcolor=blue,urlcolor=blue]{hyperref}
\usepackage{dsfont}
\usepackage{multirow}
\usepackage{mathrsfs}

\usepackage[customcolors,shade]{hf-tikz}
\graphicspath{ {./images/} }

\usepackage{colortbl}
\usepackage{changes}

\newcommand{\ket}[1]{\ensuremath{| #1 \rangle}}
\newcommand{\bra}[1]{\ensuremath{\langle #1 |}}
\newcommand{\eref}[1]{Eq.~(\ref{#1})}

\usepackage{enumitem} 

\begin{document}

\title{Strontium Ferrite Under Pressure: Potential Analogue to Strontium Ruthenate}
\date{\today}

\author{Azin Kazemi-Moridani}
\email{mohaddeseh.kazemi.moridani@umontreal.ca}
\affiliation{D\'epartement de Physique, Universit\'e de Montr\'eal, 1375 ave Th\'erèse-Lavoie-Roux, Montr\'eal, Qu\'ebec H2V 0B3, Canada}
\affiliation{Center for Computational Quantum Physics, Flatiron Institute,
162 Fifth Avenue, New York, New York 10010, USA}

\author{Sophie Beck}
\affiliation{Center for Computational Quantum Physics, Flatiron Institute,
162 Fifth Avenue, New York, New York 10010, USA}

\author{Alexander Hampel}
\affiliation{Center for Computational Quantum Physics, Flatiron Institute,
162 Fifth Avenue, New York, New York 10010, USA}

\author{A.-M. S. Tremblay}
\email{andre-marie.tremblay@usherbrooke.ca}
\affiliation{Département de Physique, Institut quantique, Université de Sherbrooke, Sherbrooke, Québec J1K 2R1, Canada}

\author{Michel Côté}
\email{Michel.Cote@umontreal.ca}
\affiliation{D\'epartement de Physique, Universit\'e de Montr\'eal, 1375 ave Th\'erèse-Lavoie-Roux, Montr\'eal, Qu\'ebec H2V 0B3, Canada}

\author{Olivier Gingras}
\email{ogingras@flatironinstitute.org}
\affiliation{Center for Computational Quantum Physics, Flatiron Institute,
162 Fifth Avenue, New York, New York 10010, USA}

\begin{abstract}
       Despite the significant attention it has garnered over the last thirty years, the paradigmatic material strontium ruthenate remains the focus of critical questions regarding strongly correlated materials.
       As an alternative platform to unravel some of its perplexing characteristics, we propose to study the isostructural and more correlated material strontium ferrite.
       Using density functional theory combined with dynamical mean-field theory, we attribute the experimentally observed insulating behavior at zero pressure to strong local electronic correlations generated by Mott and Hund's physics.
       At high pressure, our simulations reproduce the reported insulator-to-metal transition around $18$~GPa.
       Along with distinctive features of a Hund's metal, the resulting metallic state is found to display an electronic structure analogous to that of strontium ruthenate, suggesting that it could exhibit similar low-energy properties.
\end{abstract}

\maketitle

The unconventional superconductivity of strontium ruthenate (Sr$_2$RuO$_4$, SRO) still fuels debates almost thirty years after its discovery~\cite{Maeno1994, PhysRevLett.80.161, Mackenzie2017}.
It was the first layered perovskite superconductor to be discovered after the cuprates.
However, contrarily to the cuprates, SRO does not necessitate doping to exhibit superconductivity, which allows for investigations in high-quality single crystals.
This distinction has motivated extensive studies aimed at characterizing both its normal and superconducting states.

Theoretically, the normal state is nowadays understood as a correlated Hund's metal~\cite{doi:10.1146/annurev-conmatphys-020911-125045, PhysRevLett.107.256401} with important spin-orbit coupling.
Only the $t_{2g}$ electrons of the ruthenium atom play a fundamental role and interactions can be considered local, modelled by the Kanamori Hamiltonian~\cite{Fujimori1995, Zingl2019, PhysRevLett.125.166401}.
Indeed, the combination of density functional theory (DFT) and dynamical mean-field theory (DMFT) has yielded impressive agreement with experiments, reproducing for example the Fermi surface~\cite{tamai_high-resolution_2019} and the magnetic susceptibility~\cite{strand_magnetic_2019}. Additionally, it captures expected hallmarks of Hund's metals such as orbital selective mass renormalizations~\cite{bergemann_normal_2001, mravlje_coherence-incoherence_2011} and a crossover from a bad metal to a Fermi liquid~\cite{hussey_normal-state_1998, kugler_strongly_2020}.
The superconducting state, however, remains enigmatic.
The debates persist because thermodynamic measurements supported by theory suggest a one-component order parameter~\cite{PhysRevX.7.011032, Li2022, doi:10.1073/pnas.2020492118, gingras_superconducting_2019, PhysRevB.106.064513, hauck2023competition}, while other experiments observed evidence of a two-component order parameter~\cite{Benhabib2021, Ghosh2021, Grinenko2021}.
New knobs to turn could help unravel key additional information regarding SRO.

One such knob is simply to study a different, yet similar material.
In this regard, our focus turns to strontium ferrite (Sr$_2$FeO$_4$, SFO)
for which the ruthenium atom (Ru) is replaced with an isoelectronic iron atom (Fe).
This substitution results in an increased on-site Coulomb repulsion due to the more localized nature of Fe's $3d$-shell compared to Ru's $4d$-shell, along with a decreased spin-orbit coupling due to the smaller nuclear charge of Fe compared with that of Ru.
Our study of SFO is driven by a dual purpose: first, to investigate the distinctive behaviors and electronic properties exhibited by a material with an identical crystal structure to SRO, and second, to harness SFO as a potential source of deeper insights into the elusive physics of SRO's superconducting state.
This strategy has been previously successful to shed light on Hund's physics and the role of van Hove singularities by comparing SRO to Sr$_2$MoO$_4$~\cite{PhysRevLett.125.166401}.

While only a few experiments have been performed, SFO has hardly been studied and in particular no electronic structure calculation has been reported to our knowledge.
Experiments report that SFO is an antiferromagnetic insulator with a Néel temperature around $60$~Kelvin~\cite{dann_synthesis_1991, dann_structure_1993,adler_properties_1994}.
Also, a room-temperature insulator-to-metal transition has been detected around $18$~GPa~\cite{hearne_selected_1996,rozenberg_experimental_1998}.
Thus, applying pressure to SFO could be a way to suppress the antiferromagnetic order for the benefit of superconductivity, as is observed in many unconventional superconductors~\cite{Brusetti_Ribault_Jérome_Bechgaard_1982,mathur1998magnetically,lefebvre2000mott,Taillefer_Bourbonnais:2009}.

\begin{figure*}
    \includegraphics[width=0.9\textwidth]{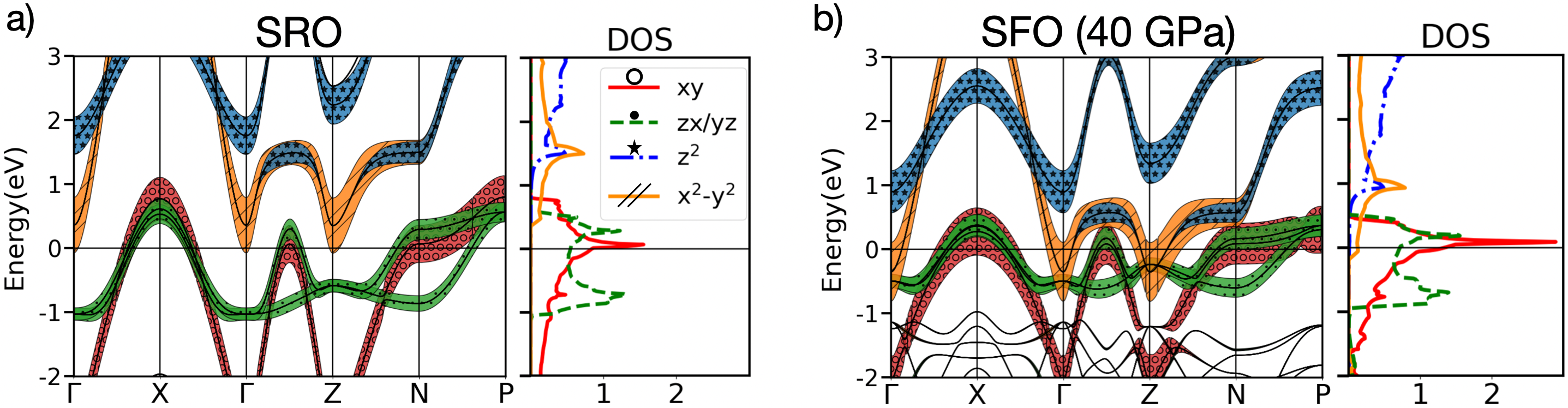}
    \caption{\label{fig:compare}
        Comparison of the open $d$-shell orbital character on the band structure of (a) SRO, and (b) SFO under $40$~GPa of isotropic pressure.
        The $d_{xy}$, $d_{yz/zx}$, $d_{z^2}$ and $d_{x^2-y^2}$ orbital characters are shown in red, green, blue and orange, respectively.
        The horizontal line at zero marks the Fermi energy.
        The $e_{g}$ orbitals are unoccupied in SRO while the $d_{x^2-y^2}$ orbital is slightly metallic in SFO at $40$~GPa.
    }
\end{figure*}

In this paper, we explore the correlated electronic structure of unstrained and strained SFO in its normal state above the Néel temperature and compare it to experiments.
Starting with DFT, we find that the electronic structure of SFO differs from that of SRO.
In SFO, both the $e_{g}$ and the $t_{2g}$ orbitals cross the Fermi energy and are partially occupied, whereas in SRO the $e_{g}$ states are empty while the $t_{2g}$ orbitals are partially occupied.
Then, by incorporating dynamical local correlations within DMFT, we explore the rich phase diagram generated by the on-site Coulomb repulsion $U$ and Hund's coupling $J$.
We argue that the phase most consistent with experiment is found around $U\geq2.5$~eV and $J<0.7$~eV. This value of $U$ is slightly above the one predicted using the constrained random phase approximation (cRPA).
In this phase, the $e_{g}$ states are pushed above the Fermi energy, while the remaining electrons in the $t_{2g}$ shell become Mott insulating.
We show that this phase undergoes an insulator-metal transition around $18$~GPa of isotropic pressure, consistent with experiments.
By comparing the band structure, the Fermi surface and the mass enhancements of this metallic phase with that of SRO, we reveal an exciting similarity between the two, suggesting SFO as an alternative platform to understand SRO.

\paragraph*{DFT electronic structure. ---}
SFO (SRO) crystallizes in a body-centered tetragonal structure with Fe (Ru) at the center of FeO$_6$ (RuO$_6$) octahedra.
The crystal field generated by the $p$-orbitals of the surrounding oxygen atoms splits the five-fold degeneracy of the Fe $d$-shell into an $e_g$ doublet ($d_{x^2-y^2}$ and $d_{z^2}$ orbitals) and a $t_{2g}$ triplet ($d_{xy}$, $d_{zx}$ and $d_{yz}$ orbitals).

Figure~\ref{fig:compare} presents the band structures of both (a) SRO and (b) SFO at 40~GPa obtained using DFT. The details of the calculations can be found in the Supplemental Materials (SM)~\cite{noteSM}.
We show the projection of the wave function onto the $d$-orbitals of the transition metal element, along with the orbital selective densities of states (DOS).
Note that the electronic structure of SFO without pressure is qualitatively similar to the one at 40~GPa~\cite{noteSM}.
In SRO, the band dispersion reveals an overlap between the $e_g$ and $t_{2g}$ orbitals, but only the $t_{2g}$ orbitals are partially filled and cross the Fermi level while the $e_g$ orbitals remain completely unoccupied.
Thus, as was done in most theoretical studies of SRO~\cite{Fujimori1995, Zingl2019, PhysRevLett.125.166401, mravlje_coherence-incoherence_2011, kugler_strongly_2020, tamai_high-resolution_2019, strand_magnetic_2019, gingras_superconducting_2019, PhysRevB.106.064513, hauck2023competition}, one can focus solely on the $t_{2g}$ orbitals.

However, in the case of SFO, both the $t_{2g}$ and $d_{x^2-y^2}$ orbitals are active at the Fermi level, necessitating a minimal model that includes the $e_g$ orbitals to describe the low-energy physics accurately.

In short, although the non-interacting band structure of SFO is similar to SRO's, the presence of $e_g$ electrons at the Fermi level is a massive distinction.
Moreover, we have been neglecting so far the role of strong electronic correlations.
In SRO, although important, they do not significantly affect the Fermi surface itself~\cite{PhysRevLett.116.106402}.
In contrast, experiments on SFO observe an insulating state rather than a metallic one.
We now investigate whether the correlation effects among the Fe $d$-electrons can be responsible for this discrepancy with DFT.

\paragraph*{Strong correlations. ---}
Because of the localized nature of $3d$ orbitals, SFO is expected to be affected by strong electronic correlations. This is reinforced by the disagreement between the \textit{ab initio} prediction of a metallic state and the experimental observation of an insulating state.
We now incorporate the missing local electronic correlations from DFT using DMFT.
This is done by projecting the DFT Kohn-Sham wave function onto a downfolded model considering only the five $3d$ orbitals of the Fe atom and constructed using the Wannier90 package.
The correlations are obtained by iteratively solving the impurity model using DMFT, with the interactions modelled by the full rotationally invariant Slater Hamiltonian (including non-density-density terms)  which depends on two parameters: the strength of the electronic Coulomb repulsion $U$ and the Hund's coupling $J$.
Details about the downfolding, the numerical calculations and the Slater Hamiltonian parametrization can be found in the SM~\cite{noteSM}.

\begin{figure}
    \includegraphics[width=\linewidth]{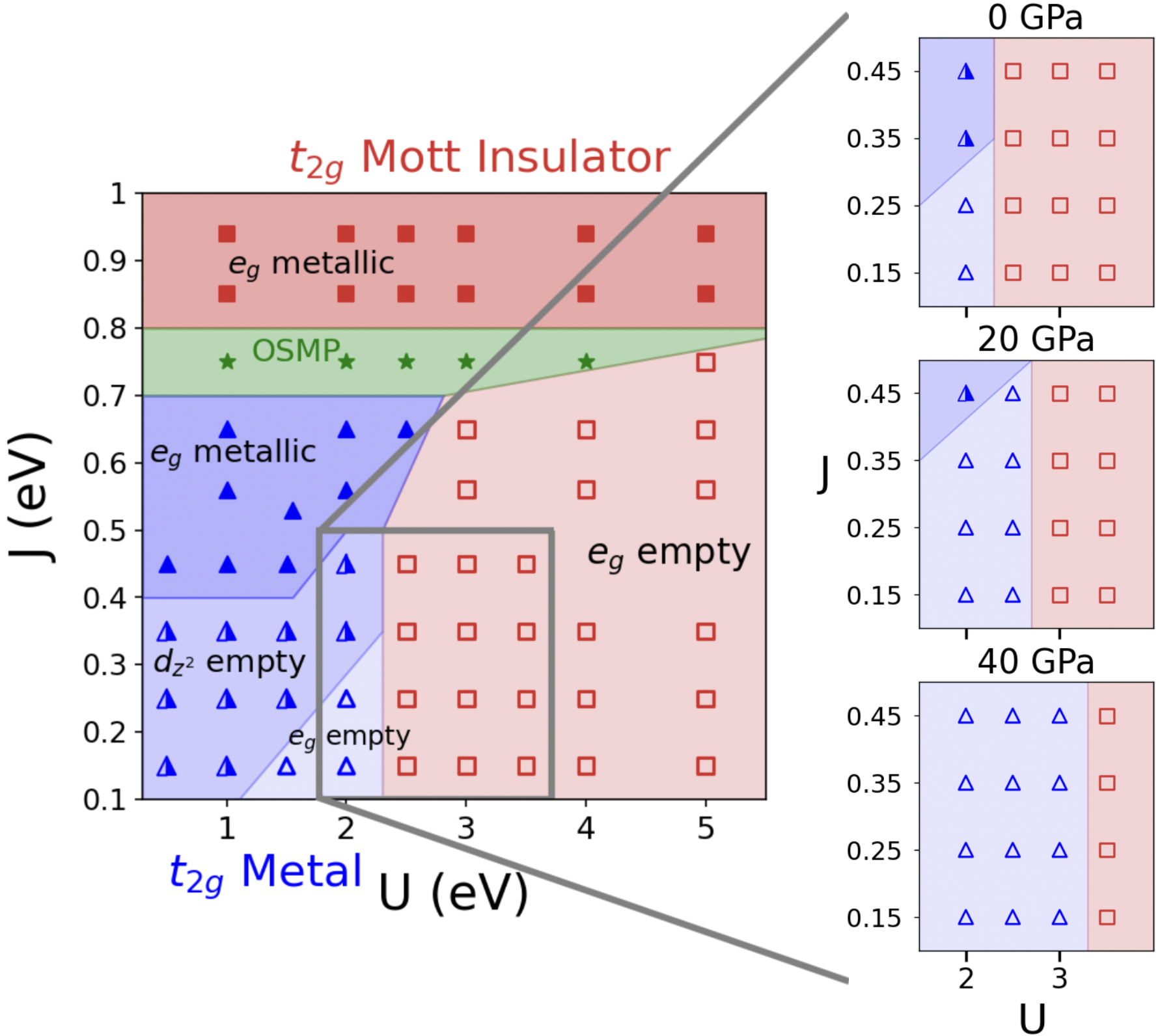}
    \caption{\label{fig:phase-diagram}
        Phase diagrams of SFO in the space of the interaction parameters at $T=146$~K for three pressures.
        Here, the Hund's coupling $J$ and on-site Coulomb repulsion $U$ are expressed in the Slater definition.
        The red squares distinguish Mott insulating $t_{2g}$ orbitals, whereas the blue triangles correspond to metallic $t_{2g}$ orbitals.
        The filling of the markers reflects whether both $e_g$ orbitals are partially occupied (full), only the $d_{x^2-y^2}$ is partially occupied (half-filled), or none are (empty).
        The narrow region with green stars corresponds to an orbital-selective Mott phase (OSMP) with $d_{xy}$ and $e_g$ metallic, and $d_{zx/yz}$ Mott insulating.
        On the right, a selected region is compared for three different pressures: $0$, $20$, and $40$~GPa.
        It highlights the insulator-to-metal transition observed around $18$~GPa~\cite{hearne_selected_1996,rozenberg_experimental_1998} for $U\sim \mathrm{2.5~eV}$, where the resulting metallic states have empty $e_{g}$ orbitals.
    }
\end{figure}

We explore possible electronic states of SFO by investigating the $U-J$ parameter space of the full Slater Hamiltonian.
The phase diagram in Fig.~\ref{fig:phase-diagram} summarizes our findings for a temperature of $146$~K ($1/k_BT = 80$~eV$^{-1}$).
Based on observables such as the spectral function~\cite{noteSM} and the resulting orbital occupations, we classify the phases using three types of colored markers: the blue triangles, red squares and green stars, corresponding to the t$_{2g}$ orbitals being metallic, being insulating due to correlations, or being orbital-selectively insulating, respectively.
For the first two classes (triangles and squares), a filled (empty) symbol represents metallic (band insulating) $e_g$ orbitals, while a half-filled symbol indicates that only the $d_{x^2-y^2}$ orbital is metallic.
For the third class (stars), the $d_{xy}$ and $e_g$ orbitals are found metallic, while the $d_{zx/yz}$ ones are Mott insulating.
We call this phase the orbital-selective Mott phase (OSMP).

We now discuss the different phases and physical mechanisms leading to the phase diagram shown in Fig.~\ref{fig:phase-diagram}. Additional information can be found in the SM~\cite{noteSM}.
In the low $U$ and low $J$ regime depicted by half-filled blue triangles, we find the DFT solution shown in Fig.~\ref{fig:compare}~(b) where only the $d_{z^2}$ orbital is empty.
As mentioned before, experimental observations suggest SFO to be a small gap insulator at zero pressure~\cite{dann_structure_1993, dann_synthesis_1991, adler_properties_1994}.
We find the phase that best reproduces these observations at larger $U$: the phase marked by open red squares where all orbitals are insulating.
This phase emerges by increasing the cost of double occupancy $U$ because it suppresses charge fluctuations and constrains the Fe atoms to host four localized electrons.
Due to the crystal field splitting generated by the oxygen atoms surrounding the Fe atom, the $d_{xy}$ orbital has the lowest on-site energy and is getting fully filled, the $e_g$ orbitals have the highest on-site energies and are pushed above the Fermi level making them band insulating, and the $d_{yz/zx}$ orbitals have to share two electrons which make them Mott insulating.

This phase, most consistent with experimental observations at zero pressure, is found roughly in the parameter regime $U\geq \mathrm{2.5}$~eV and $J<0.7$~eV.
Using the cRPA to calculate the screened interaction parameters~\cite{noteSM}, we find the static values (zero frequency limit) to be $U_{\text{cRPA}}, J_{\text{cRPA}} = 1.5$~eV, $0.5$~eV.
Although these numbers are outside of the region deemed realistic, it is known that cRPA overestimates screening effects, leading to underestimated $U$ values~\cite{Casula:2012:dyn, Honerkamp2018}.
Considering this fact, $U_{\text{cRPA}}$ appears reasonably close to the empty red square region.
Now, to attain a deeper understanding of the physical mechanisms at play and guide possible fine tuning, we continue analysing the full phase diagram.

If again we start from the small $U$ and small $J$ region, but this time go along the direction of increasing $J$ instead of $U$, we see that the occupancies of the $e_g$ orbitals start to increase. This happens because of the Hund's rule, which states that $J$ favors spin-alignment and thus spreads the orbital occupation throughout the entire $d$-shell, making all orbitals metallic at some point.
Eventually at very large $J$, there is enough occupation transfer from the $t_{2g}$ to the $e_g$ orbitals so that a Mott gap opens up in $t_{2g}$ while $e_g$ remains metallic:
first, the Mott gap opens in the less occupied $d_{yz/zx}$ orbital (leading to the green star phase), and then in the $d_{xy}$ orbital, resulting in the $t_{2g}$ insulating and $e_g$ metallic phase (full red square phase).

Finally, another remarkable result from the phase diagram of Fig.~\ref{fig:phase-diagram} is the empty blue triangle phase, which has band-insulating $e_g$ and partially filled $t_{2g}$ orbitals.
This configuration is analogous to that of SRO and could offer an alternative route to study the physics of this important system.
Since this metallic phase is on the border with the realistic insulating phase, we believe isotropic pressure might actually allow to realize this metallic phase.

\paragraph*{Isotropic pressure. ---}
In this section, we explore the insulator-to-metal transition of SFO under pressure and show that the metallic phase can be fine-tuned to have a similar band structure and Fermi surface to that of SRO. 
The insulator-to-metal transition observed experimentally happens around $18$~GPa at room temperature~\cite{hearne_selected_1996,rozenberg_experimental_1998}. Our results naturally predicts that this critical pressure should be temperature-dependent, which can be tested experimentally. 
To investigate this insulator-to-metal transition, we restricted our simulation to a window near the phase transition between the insulating (the empty red square phase) and the metallic phases. The right panels of Fig.~\ref{fig:phase-diagram} present this evolution for three pressures: $0$, $20$, and $40$~GPa.
We preserve the original crystalline symmetry, in agreement with experiments that confirmed this up to $30$~GPa~\cite{adler_optical_1994,hearne_selected_1996, rozenberg_experimental_1998}.
 
Increasing pressure increases the propensity of electrons to hop from site to site $t$, i.e., it increases the bandwidth of the $d$-shell without significantly affecting the Coulomb repulsion $U$.
Consequently, the effective Coulomb repulsion $U/t$ decreases.
This effect results in the expansion of the metallic state within the parameter space as shown on the right of Fig.~\ref{fig:phase-diagram}, while also providing a clear explanation for the insulator-to-metal transition observed in experiments.
Moreover, this effect suggests that the boundary between the metallic and insulating regions should move with temperature, leading to a temperature-dependent critical pressure for the insulator-to-metal transition. 
Experiments could already be performed to test this prediction.

We now focus on the region where both $e_{g}$ orbitals are empty, which we observe also grows with pressure. This is likely due to the modified competition between the crystal field and Hund's coupling $J$.
Indeed, applying pressure on the material increases the crystal field splitting which favors low-spin states and pushes the $e_g$ orbitals further away in energy.
In contrast, $J$ favors a high-spin state and spreads the orbital occupations.
Thus, with increasing pressure, a larger $J$ is required to occupy the $e_g$ orbitals. Therefore the empty blue triangle region that represents a metallic phase with empty $e_g$ orbitals expands.

\begin{figure}[t]
    \includegraphics[width=\linewidth]{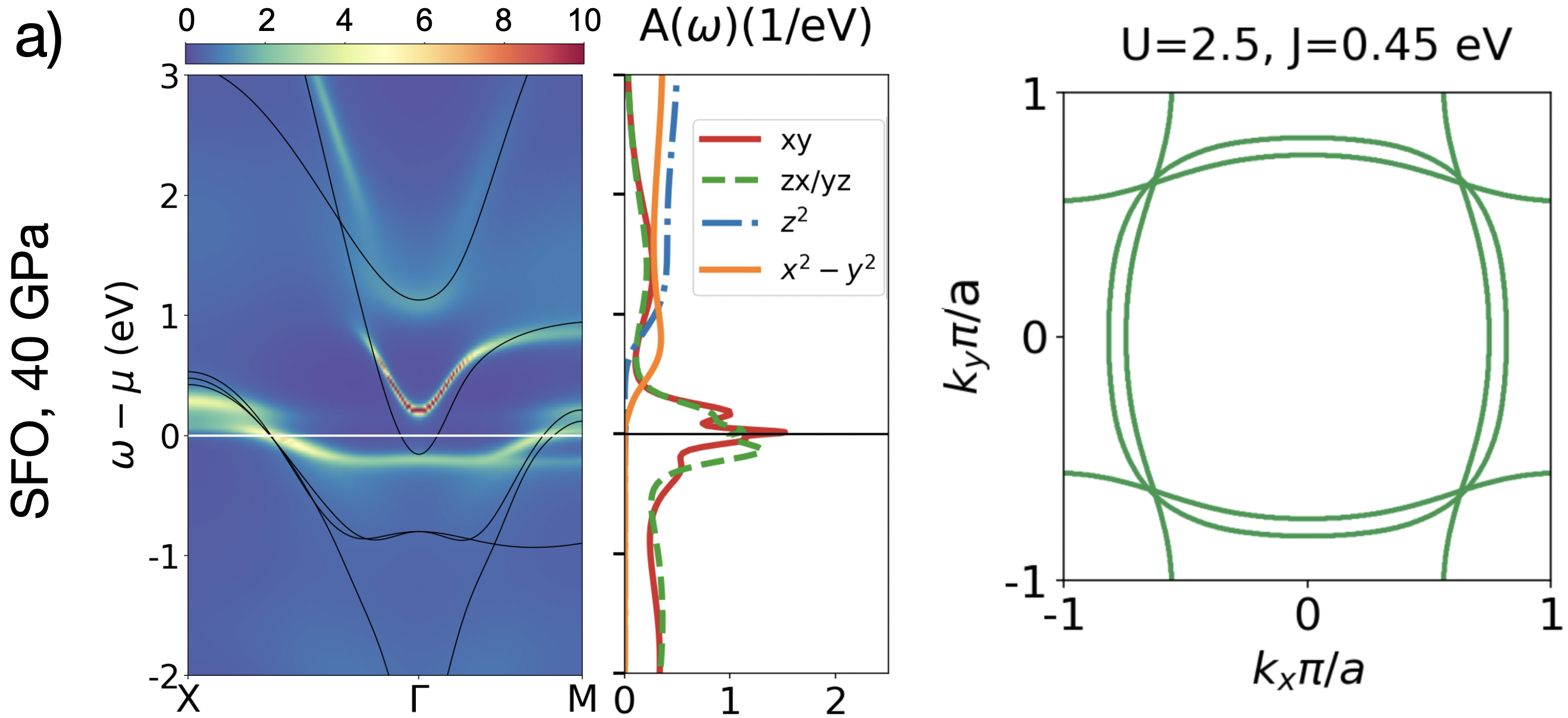}
    \includegraphics[width=\linewidth]{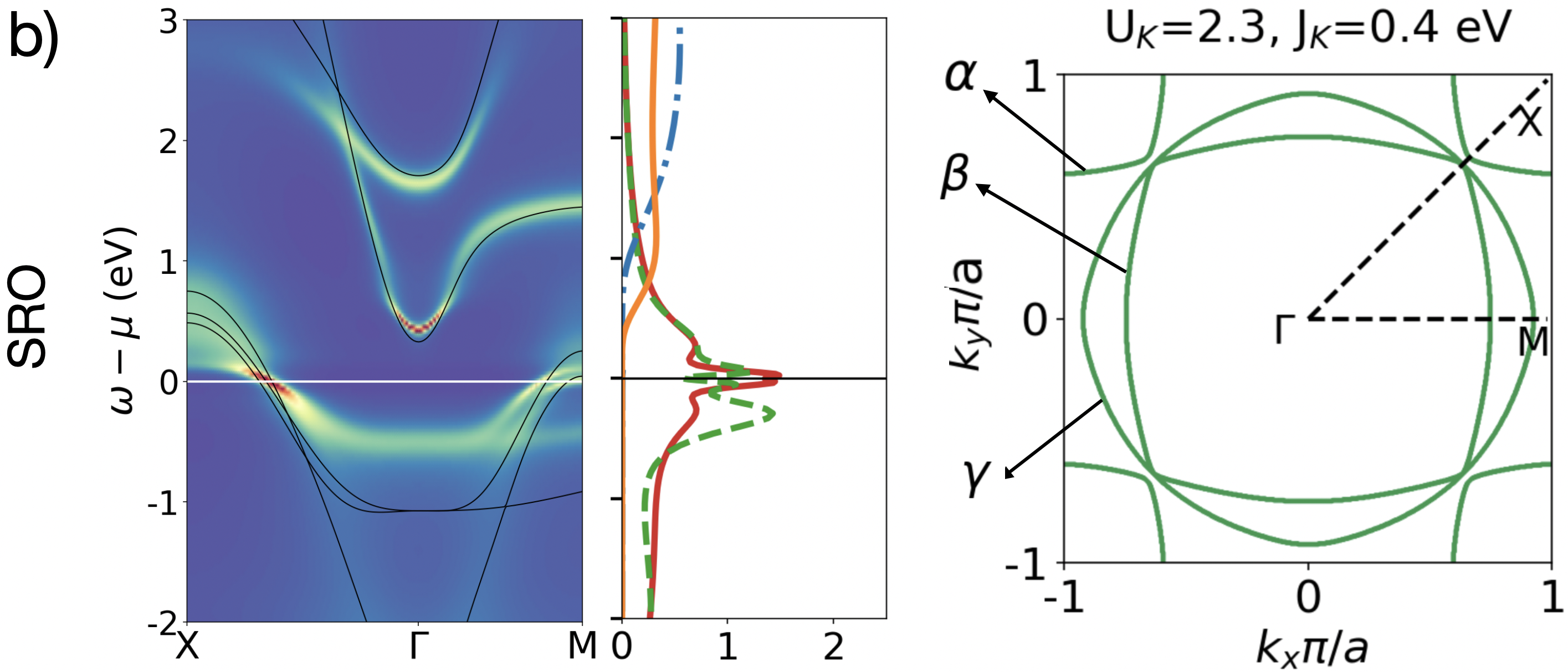}
    \caption{\label{fig:FermiS}
        Correlated band structure on the left, DOS in the middle and Fermi surface on the right of (a) $40$~GPa SFO in the three orbital metallic phase and (b) SRO, both at
        $T=146$~K. The DFT result is represented by the black lines. Clearly, correlations push $e_g$ orbitals away from the Fermi level. The Fermi surfaces are labeled $\alpha, \beta, \gamma$ in part (b). The parameters for the calculations are on top of the Fermi surfaces. Although calculations for both of the materials rely on the Slater Hamiltonian, SRO's parameters are presented in the Kanamori convention to ease the comparison with previous studies. The corresponding values of SRO in Slater are $U=1.66$~eV and $J=0.56$~eV. See Eqs~(S30, S31) of the SM for conversion relations between Slater and Kanamori values.
    } 
\end{figure} 

In order to compare the three-orbital metallic phase of SFO found under pressure with SRO, we present their respective correlated band structures, DOSs and Fermi surfaces in Fig.~\ref{fig:FermiS}.
What is meant by these \textit{correlated} objects is detailed in the SM~\cite{noteSM}.
In (a), we display the $40$~GPa phase of SFO to highlight a case at higher pressure, and we selected $U=2.5$~eV and $J=0.45$~eV as an example that reproduces the experimental observations with physically relevant parameters.
In the first panel, we contrast the correlated band structure with the one obtained using DFT and one clearly sees that correlations have pushed the $e_g$ orbitals away from the Fermi level compared to Fig.~\ref{fig:compare}~(b).

Comparing the corresponding quantities for both systems, we argue that SFO in this particular phase is analogous to SRO.
Both are metals with four $t_{2g}$ electrons, three similar Fermi sheets and comparable DOSs with a van Hove singularity in the vicinity of the Fermi energy.
There are, however, two important differences between the two, which can be regarded as opportunities:
First, even with pressure, the bandwidth of SFO remains smaller, implying stronger electronic correlations than in SRO.
Since higher pressure should bring it to a value similar to that of SRO, this represents an opportunity to study continuously a more correlated version of SRO.
This increased strength of interaction should lead to stronger magnetic fluctuations which can promote a magnetic order, or possibly superconductivity.

Second, the $\gamma$ sheet of the Fermi surface of SFO is more square-like than that of SRO.
While the calculations presented here do not include spin-orbit coupling, it should not have an important impact on SFO because of the small charge of Fe's nuclei.
As a result, the squareness of the $\gamma$ sheet presented for SFO in Fig.~\ref{fig:FermiS}~(a) should remain similar, leading to a larger nesting than in SRO. Nesting itself leads to an increased strength of the spin fluctuations.
More studies need to be performed on these speculations.

In addition to the observables presented above, we demonstrate that the three-orbital metallic phase that we find for SFO displays distinctive features of Hund's metals~\cite{doi:10.1146/annurev-conmatphys-020911-125045, PhysRevLett.107.256401, mravlje_coherence-incoherence_2011}.
This is highlighted by inspecting the effect of Hund's coupling $J$ on the orbital-selective effective mass enhancements $\frac{m^*}{m_{\text{DFT}}}\Big\lvert_l$ and on the scattering rates $\Gamma_l$.
These quantities measure the degree of electronic correlations missing from DFT and captured by DMFT. They are reported in the SM~\cite{noteSM}.
Indeed, three points stand out:
First, we find that the mass enhancements and the scattering rates all increase with $J$.
Second, the effective mass of the $xy$ orbital increases faster than those of the $yz/zx$ orbitals.
Third, the larger $J$, the lower we have to go in temperature before the effective masses saturate.
This last point highlights that it is increasingly challenging to reach the coherent regime where quasi-particles are well defined, that is the Fermi liquid regime.
The considerable increase in correlation, orbital differentiation due to $J$ and pushing of the Fermi liquid scale to lower temperatures due to $J$ are all hallmarks of Hund's metals~\cite{doi:10.1146/annurev-conmatphys-020911-125045, PhysRevLett.107.256401}.
They are also observed in SRO~\cite{mravlje_coherence-incoherence_2011}, thus supporting further the analogy between SFO and SRO.
We note that reaching the coherent regime at large $J$ is especially challenging for five orbital systems, thus we plan on extracting the effective masses that would be measured experimentally in future works.

\paragraph*{Conclusion. ---}
We studied the correlated electronic structure of strontium ferrite, Sr$_2$FeO$_4$, using the combination of density functional theory and dynamical mean-field theory.  
Correctly capturing correlation effects of the Fe $d$-electrons is essential to reproduce the experimentally observed insulating state of Sr$_2$FeO$_4$.
We find such a state for interaction strengths $U>2.5$~eV, where only the $t_{2g}$ orbitals are occupied. Moreover, we are able to reproduce the experimentally observed insulator-to-metal transition in Sr$_2$FeO$_4$ under pressure. The metallic state of Sr$_2$FeO$_4$ at $40$~GPa with $U>2.5$~eV displays the distinctive features of Hund's metals and offers a promising analogue state to Sr$_2$RuO$_4$, for which correlations could be tuned with additional pressure. Indeed, both these states are metals with four electrons in their $t_{2g}$ shells with similar band structures, density of states and Fermi surfaces. The difference is that the effective mass enhancement is generally larger in SFO, and the nesting of its Fermi surface is suggestive of enhanced magnetic fluctuations that may lead to superconductivity.

\paragraph*{Acknowledgments. ---}
We are grateful for discussions with Antoine Georges, Andrew J. Millis, Olivier Parcollet and David Sénéchal. 
Support from the Canada First Research Excellence fund is acknowledged. 
The Flatiron Institute is a division of the Simons Foundation.

\pagebreak
~
\newpage

\onecolumngrid
\begin{center}

{\large\textbf{\boldmath
Supplemental Material\\ [0.5em] {\small to} \\ [0.5em]
Strontium Ferrite Under Pressure: Potential Analogue to Strontium Ruthenatg}}\\[1.5em]

Azin Kazemi-Moridani,$^{1, 2, a}$ Sophie Beck,$^2$ Alexander Hampel,$^2$ A.-M. S. Tremblay,$^{3, b}$ Michel C\^ot\'e,$^{1, c}$ and Olivier Gingras$^{2, d}$\\[0.5em]

\textit{\small
$^{1}$D\'epartement de Physique, Universit\'e de Montr\'eal, 1375 ave Th\'erèse-Lavoie-Roux, Montréal, Québec H2V 0B3, Canada\\
$^2$Center for Computational Quantum Physics, Flatiron Institute, 162 Fifth Avenue, New York, New York 10010, USA\\
$^3$Département de Physique, Institut quantique, Université de Sherbrooke, Sherbrooke, Québec J1K 2R1, Canada
}

\vspace{2em}
\end{center}

\twocolumngrid

\setcounter{equation}{0}
\renewcommand{\theequation}{S\arabic{equation}}

\setcounter{figure}{0}
\renewcommand{\thefigure}{S\arabic{figure}}

\appendix


The content of this Supplemental Material is as follows: Section one is devoted to the computational details of the calculations performed on SFO and SRO, employing density functional theory, Wannier90, dynamical mean-field theory, and constrained random-phase approximation methods. In section two, we explain the classification and potential mechanisms found in the phase diagram of SFO. The band structures of SFO obtained by DFT, both strained and unstrained, are displayed in section three. Section four derives the relationship between Slater and Kanamori Hamiltonians. Sections five provides the definitions for correlated band structure and quasi-particle Fermi surface. Finally in section six, we report the mass enhancement and scattering rate of SFO as a function of Hund's coupling.

\section{Computational details}
In this section, we present the computational details of the density functional theory (DFT)~\cite{SM_hohenberg_inhomogeneous_1964,SM_kohn_self-consistent_1965,SM_kohn_density_1996}, construction of the downfolded models using Wannier90~\cite{SM_Pizzi_2020}, dynamical mean-field theory (DMFT)~\cite{SM_RevModPhys.68.13, SM_K.Held, SM_RevModPhys.78.865} and constrained random-phase approximation (cRPA) calculations of both SFO and SRO.

\subsubsection{DFT}
We calculated the DFT electronic structures of SFO and SRO using the ABINIT package~\cite{SM_gonze_abinit_2020,SM_romero_abinit_2020} version 9.6.2.
We used the local density approximation (LDA) functional and the projector augmented-wave (PAW) pseudo-potentials~\cite{SM_blochl_projector_1994,SM_torrent_implementation_2008} version JTH v1.1 obtained from Pseudo-Dojo~\cite{SM_pseudo-jollet, SM_pseudodojo}.
The initial crystal structures were obtained from the Materials project~\cite{SM_doi:10.1063/1.4812323} in the body-centered tetragonal unit cell (space group \textit{I}4$/mmm$ \#139) and were then relaxed.
The Brillouin zone was sampled using a 8$\times$8$\times$8 Monkhorst-Pack \textit{k}-point grid with a smearing of 0.001~Ha based on Fermi-Dirac statistics.

For SFO, 
we used a wave function energy cutoff of 33~Hartrees.
At zero pressure, we kept 37 electronic bands and obtained the following lattice parameters for the relaxed structure: 
$a=b=3.44$~\AA~and $c=11.71$~\AA.
At 40~GPa, the relaxation was performed using the stress tensor functionality of ABINIT.
We kept 45 electronic bands and obtained the following lattice parameters:
$a=b=3.27$~\AA~and $c=10.98$~\AA.

For SRO, 
we used a wave function energy cutoff of 28~Hartrees.
We kept 45 electronic bands and obtained the following lattice parameters for the relaxed structure:
$a=b=3.60$~\AA~and $c=11.77$~\AA.

\begin{figure}[h]
   \includegraphics[width=\linewidth]{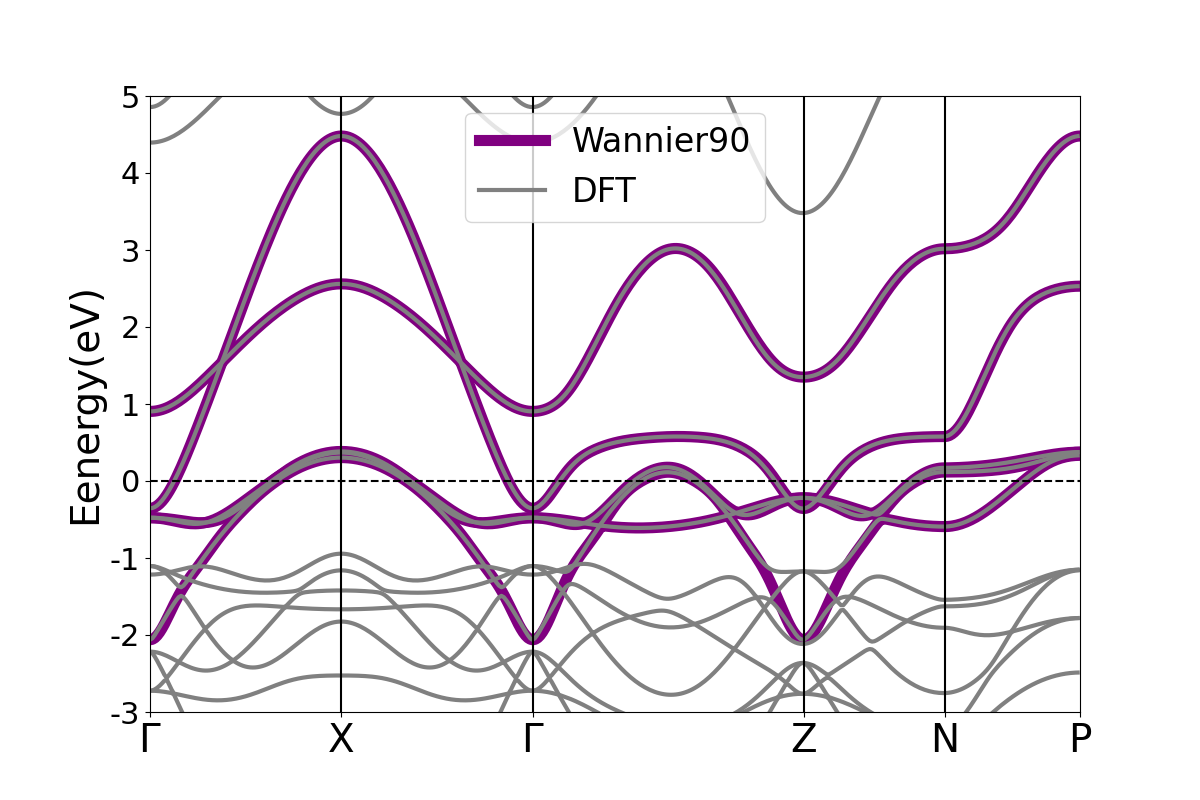}
    \caption{\label{fig:w90}
        The band structure of SFO at zero pressure obtained from Wannier90 Hamiltonian (thick line) within the localized subspace is in complete agreement with that obtained from DFT (thin line). This example serves to highlight that such agreement persists across all of our other cases.
    }
\end{figure}

\subsubsection{Wannier90}
From the Kohn-Sham wave functions obtained from DFT, the Wannier90 package allowed us to construct a downfolded model of only the active $d$ orbitals of the correlated atom, iron in SFO and ruthenium in SRO.

This is done by constructing local Wannier orbitals and maximally localizing them while still preserving the electronic dispersion of the bands near the Fermi level.
Figure~\ref{fig:w90} shows the agreement between SFO's band structure obtained from DFT and the downfolded one obtained from Wannier90.

For SFO, 
we constructed a minimal model involving all the five iron $3d$ orbitals ($t_{2g}$ and $e_g$).
At zero pressure, the Wannier orbitals were constructed within the disentanglement energy window [4.5, 12]~eV with the Fermi energy being around 6.66~eV. Under 40~GPa, the values were [5, 16]~eV, and 8.82~eV respectively.

For SRO, 
only the $t_{2g}$ orbitals are partially occupied and are needed for the minimal model.
However, to have a fair comparison to SFO with its whole $d$-shell partially occupied, we constructed the five Wannier-like $t_{2g}$ and $e_g$ orbitals of SRO in the disentanglement energy window [3, 16]~eV with the Fermi energy around 7.81~eV.

\subsubsection{DMFT}

We calculate the effects of the local interactions due to Coulomb repulsion by solving an impurity model within the DMFT framework using the TRIQS packages~\cite{SM_PARCOLLET2015398}.
The second-quantized interacting Hamiltonian responsible of these effects is expressed in a set of local orbitals with creation and annihilation operators given by $d_{o\sigma}$ where $o$ and $\sigma$ are the orbital and spin labels.
It is expressed, in the general form, as
\begin{align}
         H_{\text{int}} &= \sum_{oo'o''o'''} U_{oo'o''o'''} c^{\dag}_o c^\dag_{o'} c_{o'''} c_{o''}
\end{align}
were the matrix elements $U_{oo'o''o'''}$ are explicitly written in Eqs~(\ref{eq:def_U}, \ref{eq:def_J}, \ref{eq:coulomb_U}, \ref{eq:coulomb_J}).

In the Kanamori formulation which typically used for systems with only degenerate $t_{2g}$ orbitals, the interacting Kanamori Hamiltonian reads~\cite{SM_kanamori}
\begin{align}
    \label{eq:kanamori_ham}
         H_{\text{int}, K} &= U_K \sum_o \hat{n}_{o\uparrow} \hat{n}_{o\downarrow} + U'_K \sum_{o\neq o'} \hat{n}_{o\uparrow} \hat{n}_{o'\downarrow} \nonumber \\ 
         &+ (U'_K-J_K) \sum_{o<o',\sigma} \hat{n}_{o\sigma} \hat{n}_{o'\sigma}\\
         & - J_K \sum_{o\neq o'} d^\dag_{o\uparrow} d_{o\downarrow} d^\dag_{o'\downarrow} d_{o'\uparrow} + J_K \sum_{o\neq o'} d^\dag_{o\uparrow} d^\dag_{o\downarrow} d_{o'\downarrow} d_{o'\uparrow} \nonumber
\end{align}
where $U_K$ is the intra-orbital Coulomb repulsion term, $J_K$ is the Hund's coupling and $U'_K$ is the inter-orbital inter-spin Coulomb repulsion, usually given by $U'_K = U_K - 2J_K$ in the rotationnally invariant formulation.
The definition of the Slater $U$ and $J$ and their relation with $U_K$ and $J_K$ can be found around Eqs~(\ref{eq:conversion_U}, \ref{eq:conversion_J}).

Now the non-interacting Hamiltonian depends on the choice of these local orbitals.
The construction of these orbitals is detailed in the previous section about Wannier90.
The interface between Wannier90 and TRIQS is done using DFTTools~\cite{SM_AICHHORN2016200} and the impurity is solved using 
the continuous-time quantum Monte Carlo in the hybridization expansion formulation~\cite{SM_PhysRevLett.97.076405, SM_GULL20111078} with the CT-HYB solver~\cite{SM_SETH2016274}.
The CT-HYB solver computes the Green's function in the domain of imaginary time. The Green's function can be represented in a more compact basis by transforming it into the Legendre basis~\cite{SM_PhysRevB.84.075145}.
We used the solid\_dmft wrapper~\cite{SM_Merkel2022} to launch these calculations.
Rather than performing full charge-self-consistent DFT+DMFT calculations, 
we performed so-called "one-shot" DFT+DMFT calculations.
In this case, DMFT is used simply as a post-processing tool within solid\_dmft to solve the Hubbard-like Hamiltonian.  
To obtain the real frequency properties of the materials such as the density of states, we use the TRIQS application MaxEnt~\cite{SM_PhysRevB.96.155128} to perform the analytic continuation of the Green's functions and self-energies~\cite{SM_PhysRevB.41.2380, SM_PhysRevB.44.6011, SM_JARRELL1996133}.

\subsubsection{cRPA}

To determine the appropriate interaction parameters for DMFT, we used the constrained random phase approximation (cRPA)~\cite{SM_Aryasetiawan2004} as implemented in the RESPACK code~\cite{SM_respack2021}, which allows the calculation of the effective partially screened Coulomb interaction by separating the electronic structure into a subspace near the Fermi level and the rest of the system. Formally, this means the separation of the total electronic polarizability $P = P_{\text{sub}} + P_{\text{rest}}$ where $P_{\text{sub}}$ is the polarizability for the correlated subspace and $P_{\text{rest}}$ is for the rest of the system. Then, the screened Coulomb tensor can be calculated in a local basis from the bare Coulomb interaction tensor $\bm{V}$, as $\bm{U}(\omega) = \bm{V}/[1 - \bm{V}P_{\text{rest}}(\omega)]$. We used the same well-localized correlated subspace basis for cRPA as used within DMFT, i.e. contributions to the polarizability for the target space are removed via their overlap with the Kohn-Sham states~\cite{SM_PhysRevB.83.121101} and the screened and bare Coulomb integrals are then evaluated in the maximally localized Wannier orbitals basis set of the correlated subspace treated in DMFT. Here, we limit ourselves to the static limit $\bm{U}(\omega = 0)$ of the screened interaction.

To derive an effective symmetrized interaction tensor that can be efficiently handled by the impurity solver within DMFT, we compute the spherical average of the full four-index screened-Coulomb interaction tensor $U_{ijkl}$ to obtain the $U  := F_0$ Slater parameter, and its corresponding exchange interaction parameter $J  := (F_2+F_4)/14$ assuming $F_4/F_2=0.625$. 

While our cRPA calculations give us an estimate as to the appropriate $U$ for the studied systems, these values are not guaranteed to be quantitatively accurate for all materials properties \cite{SM_Casula:2012:dyn, SM_Honerkamp2018}, and often tend to overestimate screening, i.e. resulting in too low Coulomb interaction parameters.

\section{Classification and mechanisms found in the phase diagram}
In this section, we explain how we have classified the different phases found in the phase diagram Fig.~2 of the main text.
The classification is based on the partial densities of states obtained by analytically continuing the self-energy of the converged solutions of the phase diagram.
We also present potential mechanisms that explain this phase diagram by carefully analyzing the dominant spin configurations present in each of these phases.

\begin{figure}[t]
    \includegraphics[width=\linewidth]{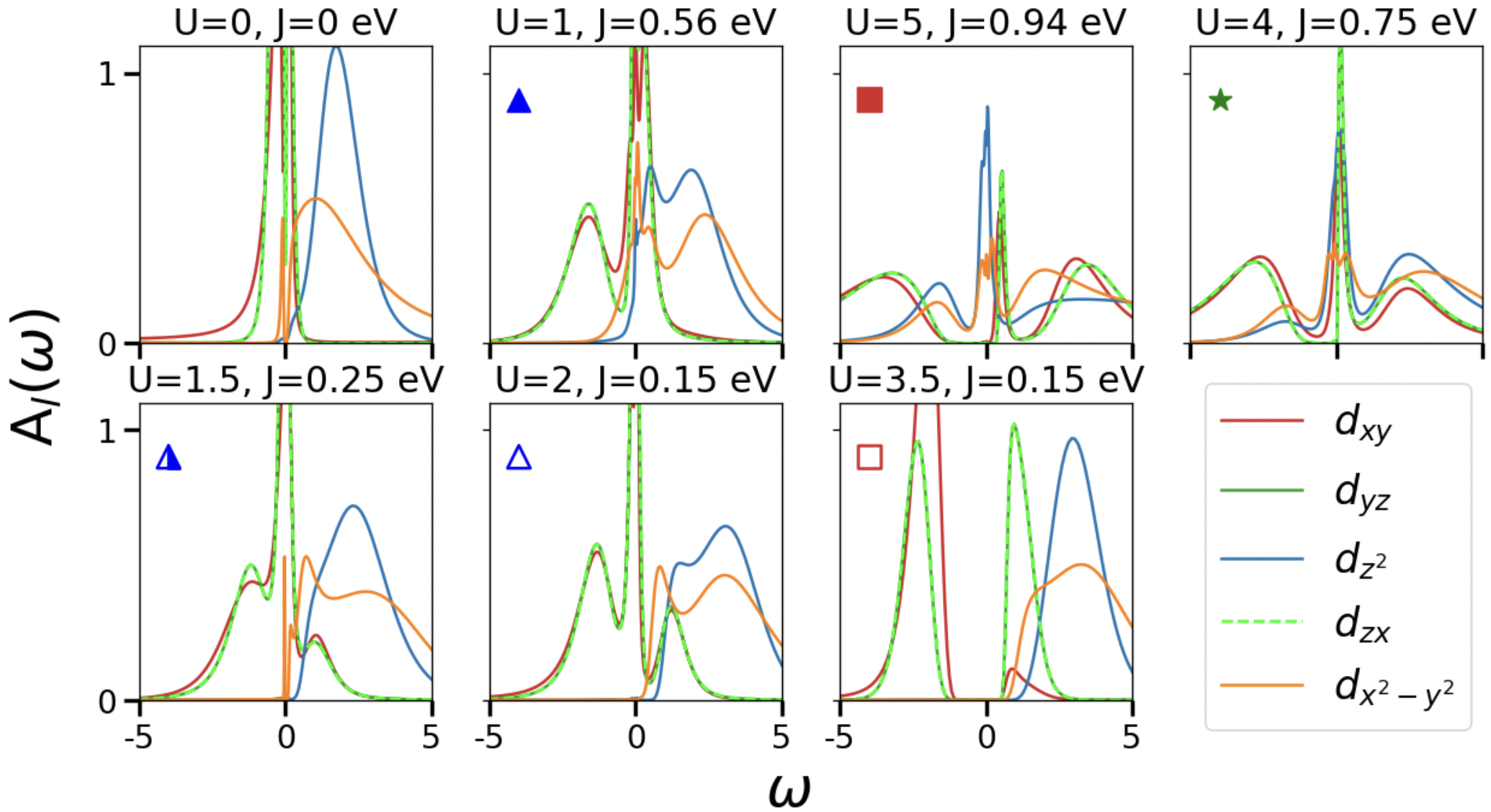}
    \caption{
        \label{fig:phase-def}
        Example of the partial density of states, or orbitally-resolved spectral functions $A_l(\omega)$, obtained in each different phase of the phase diagram in Fig.~2 of the main text.
    }
\end{figure}

\subsubsection{Density of states}
The classification of phases is done based on the orbitally-resolved spectral functions (or partial densities of states (PDOSs)) obtained at every point of the phase diagram.
These PDOSs provide information on which orbitals are metallic, band insulating or insulating due to correlations.
The orbitally-resolved spectral functions can be computed using
\begin{equation}
    A_l(\omega) = -\frac{1}{\pi} \text{Im} G_{ll}(\omega),
\end{equation}
where $G_{ll}(\omega)$ is the Green's function on the real-frequency axis corresponding to the propagation of an electron in the orbital $l$.
Since the Green's functions that we obtain from DMFT procedure are computed on the imaginary-axis, we used the MaxEnt code~\cite{SM_PhysRevB.96.155128, SM_PhysRevB.44.6011} to perform the analytic continuation of the impurity Green's function, giving as a result $G(\omega)$.

Based on these definitions, we find a total of six distinct phases in the phase diagram of SFO at zero pressure.
In Fig.~\ref{fig:phase-def}, we showcase the PDOSs of one representative point in each phase of the phase diagram.
We also show the one obtained for $U=0$, $J=0$~eV corresponding to the DFT case, although it is not explicitly included in the phase diagram.


\subsubsection{Spin configurations}
\begin{figure}[h]
    \includegraphics[width=\linewidth]{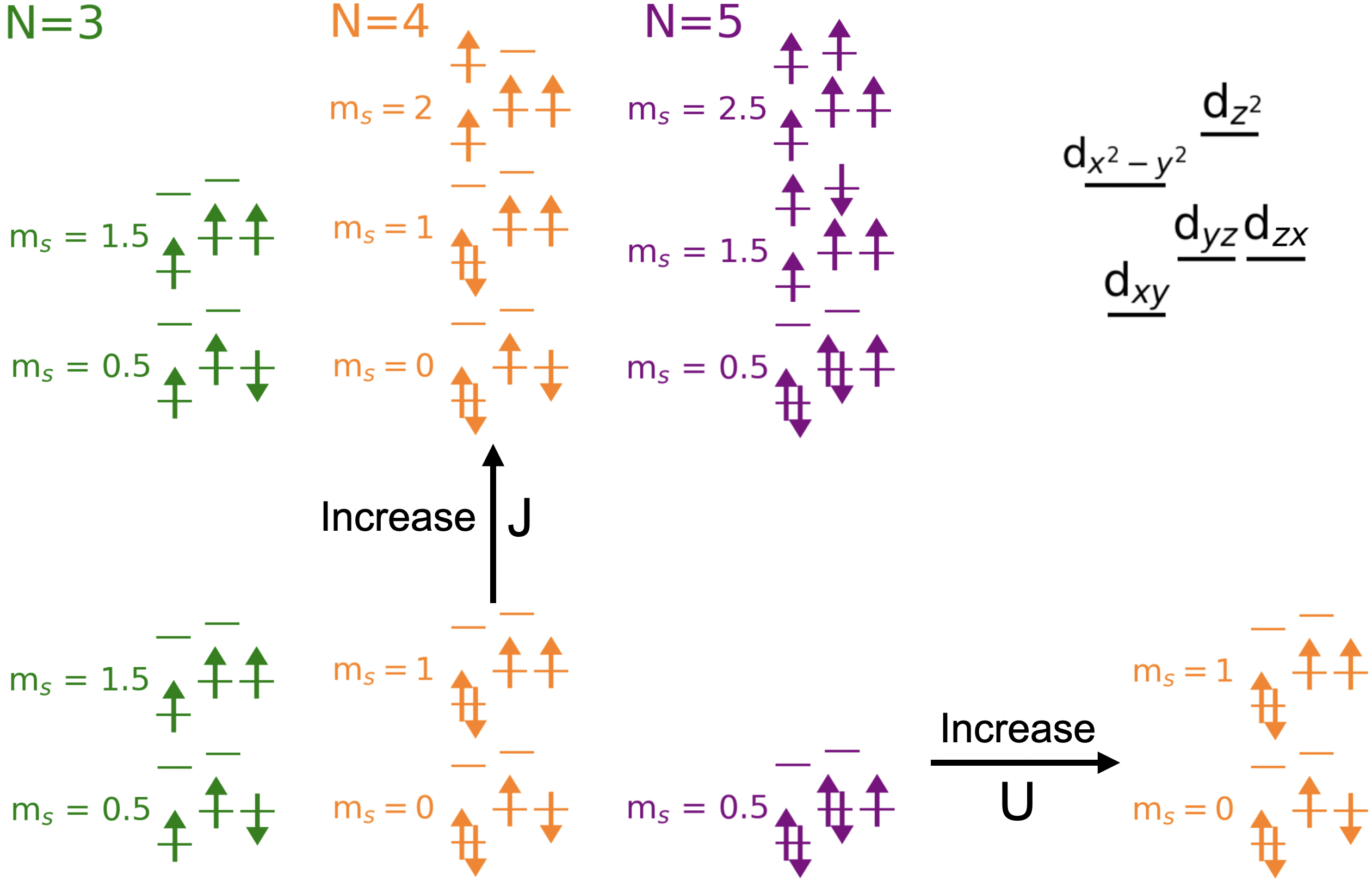}
    \caption{
        \label{fig:spin}
        Rearrangement in spin configurations of Fe $d$-electrons as a result of increasing $U$ and $J$. The strength of $U$ and $J$ goes from 0.5 to 5~eV and from 0.15 to 0.94~eV respectively. 
    }
\end{figure}

From the many-body density matrix, we extracted the eigenstates of the impurity Hamiltonian with the highest occurrence probabilities.
Since we are dealing with the Fe $d$-shell which has 10 orbitals, there are in total $2^{10}=1024$ possible states. 
We present in Fig.~\ref{fig:spin} the states with the highest probabilities at three different points in the phase diagram:
\begin{enumerate}
    \item In the low $U$ and $J$ regime corresponding to the bottom-left corner of the blue triangle phase in Fig.~2, the number of Fe electrons on the impurity can vary between $N=3$, $4$, or $5$.
    It means the electrons are able to hop around, corresponding to a metallic state.
    These three cases are distinguished in Fig.~\ref{fig:spin} with different colors: green (light grey) for $N=3$, orange (medium grey) for $N=4$, and purple (dark grey) for $N=5$. In this regime, the crystal field is dominant and mostly the $t_{2g}$ orbitals are active.
    For $N=3$, Hund's coupling favors the spreading of electrons on different orbitals.
    For larger $N$, it is preferable to first doubly filled the $xy$ orbital which has the lowest on-site energy, and then one of the $yz/zx$ orbitals.
    \item Increasing $U$ towards the bottom right corner where the empty red square phase dominates, charge fluctuations get suppressed, allowing only $N=4$ electrons on each correlated site rather than two sites to have $N=3$ and $5$ electrons.
    Again because of the strong crystal field, it is more favorable to doubly occupy the $xy$ orbital.
    \item Moving to the top of the phase diagram where $J$ has large values, the $e_{g}$ shell becomes more occupied.
    In this region, we find the Hund's metal regime, where a larger $U$ is required to drive the system to the Mott phase.
    Thus there is once again inter-site charge fluctuations leading to $N=3$ and $N=5$ states.
    However, the strong Hund's coupling now favors different total spin $|m_s|$ values.
\end{enumerate}

\begin{figure*}
    \includegraphics[width=0.9\textwidth]{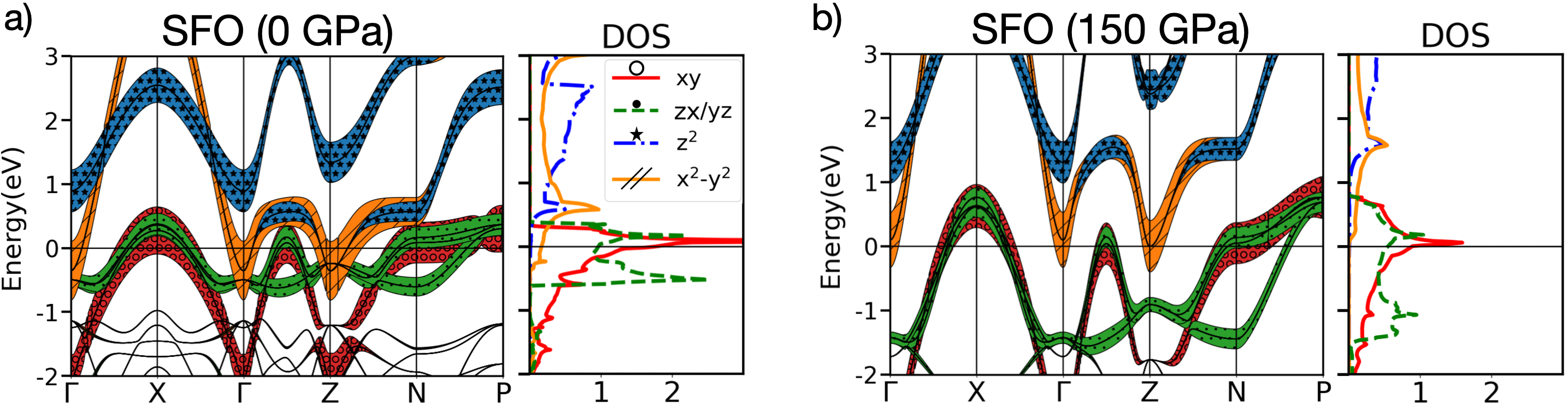}
    \caption{
        \label{fig:BS}
        Comparison of the open $d$-shell orbital character on the band structure of SFO a) under no pressure, and b) under 150~GPa of isotropic pressure.
        The $d_{xy}$, $d_{yz/zx}$, $d_{z^2}$ and $d_{x^2-y^2}$ orbital characters are shown in red, green, blue and orange, respectively.
        The horizontal line at zero marks the Fermi energy.
        The $d_{x^2-y^2}$ orbital is slightly metallic at zero pressure and almost tangent to the Fermi level at 150~GPa.
    }
\end{figure*}

\section{Band structure of Sr$_2$FeO$_4$ under pressure}

In Fig.~\ref{fig:BS}, we present the band structure of SFO a) at zero pressure on the left and b) under 150~GPa of pressure on the right.
The bandwidth of SFO under pressure becomes broader and the $e_{g}$ orbitals are pushed up in energy. However, the $d_{x^2-y^2}$ band is still touching the Fermi energy.

\section{Slater and Kanamori formulations of the local Coulomb repulsion}

In this section, we derive the effective Slater and Kanamori Hamiltonians for a specific electronic shell of orbitals, starting from the on-site Coulomb interaction $\hat{U} = \sum_{i\neq j}\frac{1}{|\textbf{r}_i-\textbf{r}_j|}$.
To do so, this interaction is first expanded in a basis of local states, as 
\begin{align}
    \label{eq:coulomb}
    \hat{U} = \frac{1}{2} \sum_{ii'jj'}\sum_{npn'p'} U^{iji'j'}_{npn'p'} d^{\dag}_{in}d^{\dag}_{jp} d_{j'p'} d_{i'n'}
\end{align}
where $d^{\dag}_{in}$ ($d_{in}$) creates (annihilates) an electron on the local site $i$ with other quantum numbers $n$.
Note the the bare Coulomb interaction does not depend on the spin of the electrons, so those are not written explicitly.
Now, the coefficients of the Coulomb interaction in that basis are given by
\begin{align}
    \label{eq:coulomb_components}
    & U^{iji'j'}_{npn'p'} = \bra{in; jp} \hat{U} \ket{i'n'; j'p'} \\
    & = \int \frac{d\textbf{r}_1 d\textbf{r}_2}{|\textbf{r}_1 - \textbf{r}_2|} \bar{\psi}_{in}(\textbf{r}_1) \bar{\psi}_{jp}(\textbf{r}_2)  \psi_{j'p'}(\textbf{r}_2) \psi_{i'n'}(\textbf{r}_1). \nonumber
\end{align}

In this derivation, we only consider the local part of this interaction where $i=i'=j=j'$. We remove these indices from now on.
Moreover, it is useful to derive the Coulomb interaction in the basis of spherical harmonics $\psi_{nlm}(\textbf{r}) = R_{nl}(r)Y_m^l(\theta, \phi)$ which solve the hydrogen atom, where $n,l,m$ are the principal, total angular momentum and projected angular momentum quantum numbers respectively.
The position is expressed in spherical coordinates $\textbf{r} = r\sin \theta\cos \phi \textbf{x} + r \sin \theta \sin \phi \textbf{y} + r\cos \theta \textbf{z}$ and $R, Y$ are the radial and angular parts of the spherical harmonics.
From now on, we only consider a single electronic shell, so this basis has fixed $n$ and $l$, and the only varying quantum number is $m$.

In this basis, we can write
\begin{equation}
    \label{eq:1_over_r_harmonics}
    \frac{1}{|\textbf{r}_1-\textbf{r}_2|} = \sum_{k=0}^\infty \frac{r^k_<}{r^{k+1}_>} \frac{4\pi}{2k+1} \sum_{q=-k}^k Y^k_q(\Omega_2) \bar{Y}^k_q(\Omega_1)
\end{equation}
where $\Omega \equiv (\theta, \phi)$ is a solid angle, $r_<$ ($r_>$) is the smaller (larger) of $r_1$ and $r_2$, and $\bar{Y}\equiv Y^*$.
We insert this expression in \eref{eq:coulomb_components} and find 
\begin{equation}
    U_{m_1m_2m_3m_4} = \sum_{k=0}^{2l} a_k(m_1m_3;m_2m_4) F_k
\end{equation}
where we defined the Slater integrals
\begin{equation}
    F_k \equiv \int dr_1 dr_2 \ r_1^2r_2^2 \ R_{nl}^2(r_1) \frac{r^k_<}{r^{k+1}_>} R^2_{nl}(r_2)
\end{equation}
and the angular integrals given by
\begin{equation}
    \frac{a_k(m_1m_2;m_3m_4)}{4\pi} \equiv \sum_{q=-k}^k \frac{G_q^k(m_1,m_2) \left[G_q^k(m_4,m_3)\right]^*}{2k+1} 
\end{equation}
with the Gaunt coefficients for $l$ defined as
\begin{equation}
    G_q^k(m,m') \equiv \int d\Omega \ \bar{Y}_m^l(\Omega) Y_q^k(\Omega) Y_{m'}^l(\Omega).
\end{equation}

The most important components of the Coulomb interaction are the direct ($U_{mm'mm'}$) and exchange ($U_{mm'm'm}$ with $m\neq m'$) integrals, which we write as
\begin{align}
    \label{eq:def_U}
    U_{mm'mm'} \equiv U_{mm'} = \sum_{k=0}^{2l} b_k(m,m')F_k \\
    \label{eq:def_J}
    U_{mm'm'm} \equiv J_{mm'} = \sum_{k=0}^{2l} c_k(m,m')F_k.
\end{align}
It can be shown that they are positive and that $U_{mm'} \geq J_{mm'}$.
In this basis, one can show that $U_{mmm'm'} \equiv K_{mm'} = U_{mm'}\delta_{mm'}$. Neglecting the other terms, the Coulomb interaction \eref{eq:coulomb} in the local approximation becomes, with $n_{m\sigma} \equiv d^\dag_{m\sigma}d_{m\sigma}$,
\begin{align}
    \hat{U}_{\text{loc}} & = \frac{1}{2} \sum_{mm'} \sum_{\sigma} U_{mm'} n_{m\sigma} n_{m'-\sigma} \\
        & + \frac{1}{2} \sum_{m\neq m'}\sum_{\sigma} \left( U_{mm'} - J_{mm'} \right) n_{m\sigma} n_{m'\sigma} \\
        & + \frac{1}{2} \sum_{m\neq m'} \sum_{\sigma} J_{mm'} d^{\dag}_{m\sigma} d^{\dag}_{m'-\sigma} d_{m-\sigma} d_{m'\sigma} \\
        & + \frac{1}{2} \sum_{m\neq m'} \sum_{\sigma} K_{mm'} d^{\dag}_{m\sigma} d^{\dag}_{m-\sigma} d_{m'-\sigma} d_{m'\sigma}.
\end{align}

We now expressed this interaction in terms of the average Coulomb parameters in the basis of spherical harmonics, defined as
\begin{align}
    U_{avg} & = \frac{1}{(2l+1)^2} \sum_{mm'} U_{mm'} \quad \text{and} \\
    U_{avg} - J_{avg} & = \frac{1}{2l(2l+1)}\sum_{mm'} \left( U_{mm'} - J_{mm'} \right).
\end{align}

Now in materials, real harmonics (that we now call orbitals) are normally used, because they are better eigenstates of the crystal fields. They are defined as follow:
\begin{align}
    y_{l0} & \equiv Y_0^l, \ 
    y_{lm} \equiv \frac{1}{\sqrt{2}}\left( Y_{-m}^l + (-1)^m Y_m^l \right) \quad \text{and} \\
     y_{l-m} & \equiv \frac{i}{\sqrt{2}}\left( Y_{-m}^l - (-1)^mY_m^l \right), \ \text{for} \ m > 0.
\end{align}
In this orbital basis, we use the letter $o$ to denote a real spherical harmonic.
We define
\begin{equation}
    \mathcal{J}_{avg} = \frac{1}{2l(2l+1)} \sum_{o\neq o'} J_{oo'}
\end{equation}
and one can show that in the $l=2$ case corresponding to $d$-orbitals,
\begin{equation}
    \mathcal{J}_{avg} = \frac{5}{7}J_{avg}.
\end{equation}
Moreover, in this basis, the terms $K_{oo'}$ are non-vanishing for off-diagonal elements. 

We now look specifically at $d$ electrons with a total angular momentum $l=2$.
The basis of real spherical harmonics (orbitals) is chosen as $\{d_{xy}, d_{yz}, d_{z^2}, d_{zx}, d_{x^2-y^2}\}$.
In this basis, the $U_{oo'}$ part of the Coulomb interaction is given as
\begin{align}
    \label{eq:coulomb_U}
    & U_{oo'} = \\
    & \left[ 
        \begin{array}{ccccc} 
            U_0 & U_0-2J_1 & U_0-2J_2 & U_0-2J_1 & U_0-2J_3 \\
            U_0-2J_1 & U_0 & U_0-2J_4 & U_0-2J_1 & U_0-2J_1 \\
            U_0-2J_2 & U_0-2J_4 & U_0 & U_0-2J_4 & U_0-2J_2 \\
            U_0-2J_1 & U_0-2J_1 & U_0-2J_4 & U_0 & U_0-2J_1 \\
            U_0-2J_3 & U_0-2J_1 & U_0-2J_2 & U_0-2J_1 & U_0
        \end{array}
    \right], \nonumber
\end{align}
while the $J_{oo'}=K_{oo'}$ parts are given as
\begin{align}
    \label{eq:coulomb_J}
    & J_{oo'} = K_{oo'} =
    \left[ 
        \begin{array}{ccccc} 
            U_0 & J_1 & J_2 & J_1 & J_3 \\
            J_1 & U_0 & J_4 & J_1 & J_1 \\
            J_2 & J_4 & U_0 & J_4 & J_2 \\
            J_1 & J_1 & J_4 & U_0 & J_1 \\
            J_3 & J_1 & J_2 & J_1 & U_0
        \end{array}
    \right].
\end{align}
In these expressions, 
\begin{align}
    U_0 & = F_0 + \frac{4}{49}F_2 + \frac{4}{49}F_4, \\
    J_1 & = \frac{3}{49}F_2 + \frac{20}{441}F_4, \\
    J_2 & = \frac{4}{49}F_2 + \frac{5}{147}F_4, \\
    J_3 & = \frac{5}{63}F_4, \\
    J_4 & = \frac{1}{49}F_2 + \frac{10}{147}F_4.
\end{align}
This representation of the local Coulomb interaction for $d$-electrons is called the Slater Hamiltonian and is parameterized by the Slater integrals $F_0$, $F_2$ and $F_4$.
The standard notation of the Slater Hamiltonian uses $U\equiv F_0$, $J\equiv (F_2+F_4)/14$ and $F_2/F_4$ is fixed at $0.625$.

Projecting only on the $t_{2g}$ orbitals ($\{d_{xy}, d_{yz}, d_{zx}\}$), we find
\begin{align}
    & U_{t_{2g}} = 
    \left[ 
        \begin{array}{ccccc} 
            U_0 & U_0-2J_1 & U_0-2J_1 \\
            U_0-2J_1 & U_0 & U_0-2J_1 \\
            U_0-2J_1 & U_0-2J_1 & U_0
        \end{array}
    \right] \nonumber
\end{align}
and 
\begin{align}
    & J_{t_{2g}} = K_{t_{2g}} = 
    \left[ 
        \begin{array}{ccccc} 
            U_0 & J_1 & J_1 \\
            J_1 & U_0 & J_1 \\
            J_1 & J_1 & U_0
        \end{array}
    \right]. \nonumber
\end{align}
In this $t_{2g}$ subspace, $\hat{U}_\lvert{t_{2g}}$ is called the Kanamori Hamiltonian and is parameterized solely by $U_0$ and $J_1$, referred to as $U_K$ and $J_K$ in the main text.
One can show that the Slater and the Kanamori parameters are related by
\begin{align}
    \label{eq:conversion_U}
    U_0 & \equiv U_K =  U + \frac{8}{7}J \quad \text{and} \\
    \label{eq:conversion_J}
    J_1 & \equiv J_K = \frac{\frac{6}{7}\frac{F_2}{F_4} + \frac{40}{63}}{1+\frac{F_2}{F_4}} J \sim \frac{5}{7} J.
\end{align}





\section{Correlated band structure}
A band structure typically presents the single-particle energy states of infinite lifetime quasiparticles as a function of momentum, obtained using band theory.
We call the \textit{correlated} generalization the plot of the lattice spectral function $A(\textbf{k},\omega)$, proportional to the density of states. 
In the non-interacting case, it boils down exactly to a typical band structure, but in the presence of interactions, the bands can be broadened by a finite quasi-particle lifetime acquired due to interactions between electrons. 

To obtain the lattice spectral function, we first have to construct the lattice Green's function because
\begin{equation}
    A(\textbf{k},\omega) = -\frac{1}{\pi} \text{Im} G(\textbf{k},\omega).
\end{equation}
The lattice Green's function is defined in the following way:
\begin{equation}
    G(\textbf{k},\omega) = \frac{1}{\omega + \mu - \epsilon(\textbf{k}) - \Delta \Sigma(\textbf{k}, \omega)}
\end{equation}
where $\mu$ is the chemical potential, $\epsilon(\textbf{k})$ is the non-interacting Hamiltonian obtained within Wannier90 and $\Delta \Sigma(\textbf{k}, \omega)$ is the lattice self-energy for which double-counting was subtracted.
This quantity is obtained by taking the impurity self-energy that was analytically continued $\Sigma_{\text{imp}}(\omega)$, removing the double-counting $\Sigma_{\text{DC}}$, and re-embedding to the lattice using the Wannier90 projectors $P_{\textbf{k} \nu l}$ that allows to project the band $\mu$ at the $\textbf{k}$ momentum to the $l$ orbital. In the band basis, the components of this self-energy are given by
\begin{equation}
    \label{eq:emb_self_energy}
    \left[\Delta\Sigma(\textbf{k}, \omega)\right]_{\nu\nu'} = \sum_{ll'} P^*_{\textbf{k}\nu l} \left[ \Sigma_{\text{imp}}(\omega) - \Sigma_{\text{DC}} \right]_{ll'} P_{\textbf{k} l' \nu'}.
\end{equation}

Performing a summation over $k$ of the spectral function gives us $A(\omega)$ which can be compared with the density of states from DFT. 
Instead of the self-energy, one can also analytically continue the impurity Green's function from the imaginary axis, $G_{\text{imp}}(\tau)$, to the real axis, $G_{\text{imp}}(\omega)$, using MaxEnt. Therefore, the spectral function $A(\omega)$ can also be obtained from $G_{\text{imp}}(\omega)$.
The orbitally-resolved spectral functions of Fig.~\ref{fig:phase-def} are obtained from analytical continuation of the impurity Green's function.

\subsection{Quasi-particle Fermi surface}

The lattice Green's function in the band basis can written as

\begin{equation}
    G_{\nu \nu'}(\textbf{k},\omega) = \frac{1}{(\omega + \mu - \epsilon_\nu(\textbf{k}))\delta_{\nu \nu'} - \Delta\Sigma_{\nu \nu'}(\textbf{k},\omega)},
\end{equation}
where $\nu \nu'$ are band indices. 

\noindent
The quasi-particle Fermi surface is the $\omega=0$ solution of the poles of the above Green's function, i.~e. when the quasi-particle dispersion relation crosses the Fermi level

\begin{equation}
    \det[(\omega +\mu - \epsilon_\nu(\textbf{k}))\delta_{\nu \nu'} - \Delta \Sigma'_{\nu \nu'}(\textbf{k},\omega)] = 0
\end{equation}
where $\Delta \Sigma'_{\nu \nu'}$ is the real part of the self-energy defined in Eq.~(\ref{eq:emb_self_energy}).
In the quasi-particle approximation, quasi-particles are presumed to have infinite lifetime and the imaginary part $\Delta\Sigma''$ is neglected.

\section{Mass enhancement}

For an  electron on the orbital $l$, the enhancement of the effective mass due to electronic correlations captured by the DMFT self-energy $\Sigma$ compared to bare one given by DFT is given, on the imaginary-axis, by
\begin{equation}
    \frac{m^{\ast}}{m_{\text{DFT}}}\Big\lvert_l=1-\frac{\partial \mathrm{Im}\Sigma_{l}(i\omega)}{\partial (i\omega)}\Big\lvert_{i\omega\rightarrow0^+},
\end{equation}
where $i\omega$ is the Matsubara frequency.
This enhancement captures the renormalization of the DFT bands due to electronic correlations.
In a non-interacting system, there is no self-energy ($\Sigma=0$) and the mass enhancement is 1, whereas in strongly correlated systems, $\frac{m^{\ast}}{m_{\text{DFT}}}$ has a value larger than one, which indicates that the quasi-particles have a heavier effective mass due to the electron-electron interactions.
Because interactions generally affect each orbital differently, $\frac{m^{\ast}}{m_{\text{DFT}}}$ is orbital-specific.

The concept of quasi-particle becomes no longer relevant when electronic states are completely filled or empty.
Therefore, we computed the mass enhancement only for the metallic $t_{2g}$ orbitals of SFO under 40~GPa.
These are obtained by fitting a fourth-order polynomial to the the imaginary part of $\Sigma_{l}(i\omega)$ on the six lowest Matsubara frequencies.
This fit allows us to extract the derivative $\frac{\partial \mathrm{Im}\Sigma_{l}(i\omega)}{\partial (i\omega)}\Big\lvert_{\omega\rightarrow0^+}$, corresponding to the effective mass enhancement due to electronic correlations. 
We can also extract the scattering rate $\Gamma_l = -Z_l\mathrm{Im}\Sigma_{l}(i\omega)\lvert_{\omega\rightarrow0^+}$ from this fit, as the intercept of the polynomial with the y-axis.

We present the mass enhancement and scattering rates in Table~\ref{tab:eff_mass_enhancement} and Table~\ref{tab:scat_rate_enhancement} respectively, at fixed $U_K=3$~eV and as a function of the inverse temperature $\beta$ and Hund's coupling $J_K$. In Table~\ref{tab:eff_mass_enhancement}, we compare with the effective masses calculated for SRO and reported in Ref.~\citenum{SM_mravlje_coherence-incoherence_2011}.

We keep $U_K$ fixed, because it corresponds to the standard cost of double occupancy in the second-quantized interacting Hamiltonian of Eq.~\ref{eq:kanamori_ham} and varying only $J_K$ highlights better the effect of Hund's coupling and makes the comparison with Ref.~\citenum{SM_mravlje_coherence-incoherence_2011} straight forward.

\begin{table*} [t]
    \centering
    \caption{
        \label{tab:eff_mass_enhancement}
        Evolution of the orbital specific effective mass enhancement due to electronic correlations as a function of Hund's coupling for both $40$~GPa SFO in the three orbital metallic phase and SRO. We compute the enhancement at $\beta = 1/k_BT = 80$~eV$^{-1}$, $100$~eV$^{-1}$ and $150$~eV$^{-1}$. We keep $U_K$ fixed at $3$~eV.
        The uncertainties smaller than $0.1$ are not specified in the table.
    }
    \begin{tabular}{c c c |c c |c c |c c}
        \multicolumn{3}{c}{SFO ($40$~GPa), $U_K=3$~eV, $\beta=80$~eV$^{-1}$} & \multicolumn{2}{c}{$\beta=100$~eV$^{-1}$} & \multicolumn{2}{c}{$\beta=150$~eV$^{-1}$} &
        \multicolumn{2}{c}{SRO, $U_K=2.3$~eV}\\ 
        \hline
        \hline
        \ $J_K$~(eV) \ & \ $\frac{m^*}{m_\text{LDA}}\Big|_{xy}$ \ & \ $\frac{m^*}{m_\text{LDA}}\Big|_{yz/zx}$ \ & \
        $\frac{m^*}{m_\text{LDA}}\Big|_{xy}$ \ & \
        $\frac{m^*}{m_\text{LDA}}\Big|_{yz/zx}$ \ & \
        $\frac{m^*}{m_\text{LDA}}\Big|_{xy}$ \ & \
        $\frac{m^*}{m_\text{LDA}}\Big|_{yz/zx}$ \ & \ 
        $\frac{m^*}{m_\text{LDA}}\Big|_{xy}$ \ & \
        $\frac{m^*}{m_\text{LDA}}\Big|_{yz/zx}$\\ 
        \hline
        0.1 & 2.9 $\pm$ 0.1 & 2.6 & 3.1 & 2.7 & 3.2 $\pm$ 0.1 & 2.5 $\pm$ 0.2 & 1.7 & 1.7\\  
        0.2 & 3.6 $\pm$ 0.1 & 3.4 & 4.2 & 3.7 & 5.2 $\pm$ 0.2 & 3.9 $\pm$ 0.1 & 2.3 & 2\\
        0.3 & 3.8 & 3.8 & 4.6 $\pm$ 0.2 & 4.2 & 6.1 $\pm$ 0.4 & 5.0 $\pm$ 0.2 & 3.2 & 2.4\\
        0.4 & 3.7 $\pm$ 0.1 & 3.6 & 4.4 & 4.2 & 6.3 $\pm$ 0.3 & 5.4 $\pm$ 0.1 & 4.5 & 3.3\\
        \hline
        \hline
    \end{tabular}
\end{table*}

\begin{table*} [t]
    \centering
    \caption{
        \label{tab:scat_rate_enhancement}
        Evolution of the orbital specific scattering rate due to electronic correlations as a function of the Hund's coupling for both $40$~GPa SFO in the three orbital metallic phase and SRO. We compute the enhancement at $\beta = 1/k_BT = 80$~eV$^{-1}$, $100$~eV$^{-1}$ and $150$~eV$^{-1}$. We keep $U_K$ fixed at $3$~eV. The uncertainties below 10~$\%$ are not explicitly specified.
    }
    \begin{tabular}{c c c |c c|c c}
        \multicolumn{3}{c}{SFO ($40$~GPa), $U_K=3$~eV, $\beta=80$~eV$^{-1}$} & \multicolumn{2}{c}{$\beta=100$~eV$^{-1}$} & \multicolumn{2}{c}{$\beta=150$~eV$^{-1}$} \\ 
        \hline
        \hline
        \ $J_K$~(eV) \ & \ $\Gamma_{xy}$~(meV) \ & \ $\Gamma_{yz/zx}$~(meV) \ & \
         $\Gamma_{xy}$~(meV) \ & \ $\Gamma_{yz/zx}$~(meV) \ & \
        $\Gamma_{xy}$~(meV) \ & \ $\Gamma_{yz/zx}$~(meV) \\ 
        \hline
        0.1 & 10.9 $\pm$ 1.2 & 4.0 $\pm$ 0.5 & 6.8 $\pm$ 0.8 & 1.0 $\pm$ 0.3 & 4.6 $\pm$ 1.1 & 2.3 $\pm$ 1.9\\  
        0.2 & 29.3 & 12.1 & 18.7 & 5.6 & 8.4 $\pm$ 1.3 & 2.2 $\pm$ 0.7\\
        0.3 & 45.9 & 19.0 & 31.2 & 11.6 & 15.5 $\pm$ 2.2 & 4.0 $\pm$ 0.8\\
        0.4 & 62.1 & 28.0 & 44.4 & 17.8 & 22.1 & 6.8\\
        \hline
        \hline
    \end{tabular}
\end{table*}

The results reported in these tables highlight characteristics that the clearly establish the metallic state of SFO under pressure as a Hund's metal.
Indeed, we observe that, similarly to what was observed in SRO~\cite{SM_mravlje_coherence-incoherence_2011}, the Hund's coupling leads to:
\begin{enumerate}[label=(\roman*)]
    \item An increase of electronic correlations apparent in the effective masses and scattering rates,
    \item An orbital selective enhancement of the effective masses of the Fe $d$-shell, and 
    \item A push of the Fermi liquid crossover to lower temperature and a broader temperature range for the incoherent regime. 
\end{enumerate}
This last point is the reason why it is more difficult to calculate the effective masses with increasing $J_K$: we need to reach lower temperature to be in the coherent regime where the effective masses saturate. This is also supported by the larger scattering rate. In our case, even at $\beta=150$~eV, the effective masses for $J_K = 0.4$~eV are far from this saturation.

\end{document}